# Uncovering Non-native Speakers' Experiences in Global Software Development Teams——A Bourdieusian Perspective

Yi Wang[1,2]*, Yang Yue[3#], Wei Wang[4], and Gaowei Zhang[4]

[1]Department of Software Engineering, Beijing University of Posts and Telecommunications, 10# Xitucheng Road, Beijing, 100876, China
[2]Key Laboratory of Trustworthy Distributed Computing and Service (MOE), 10# Xitucheng Road, Beijing, 100876, China
[3]Department of Informatics, University of California-Irvine, Irvine, CA 92612, USA.
[4]School of Artificial Intelligence, Beijing University of Posts and Telecommunications, 10# Xitucheng Road, Beijing, 100876, China

*Corresponding author(s). E-mail(s): yiwang@bupt.edu.cn;
#Current Affliation: Department of Computer Science & Information Systems, California State University-San Marcos, San Marcos, CA, 92096, USA
Contributing authors: yyue@csusm.edu; weiwang@bupt.edu.cn; zhanggaowei@bupt.edu.cn

**Abstract.** Globally distributed software development has been a mainstream paradigm in developing modern software systems. We have witnessed a fast-growing population of software developers from areas where English is not a native language in the last several decades. Given that English is still the de facto working language in most global software engineering teams, we need to gain more knowledge about the experiences of developers who are non-native English speakers. We conducted an empirical study to fill this research gap. In this study, we interviewed 27 Chinese developers in commercial software development and open source global software development teams and applied Bourdieu's capital-field-habitus framework in an abductive data analysis process. Our study reveals four types of capital (language, social, symbolic, and economic) involved in their experiences and examines the interrelations among them. We found that non-native speakers' insufficient language capital played an essential role in prohibiting them from accessing and accumulating other capital, thus reproducing the sustained and systematic disadvantaged positions of non-native English speakers in GSD teams. We further discussed the theoretical and practical implications of the study.

# 1. Introduction

Software development has been increasingly an international engineering activity (Herbsleb 2007; Herbsleb et al. 2001). An essential feature of global software development (GSD) is the involvement of a large amount of developers from multiple countries around the world (Calefato et al. 2021). Members of such a globally distributed team will likely speak different native languages. Thus, English often serves the Lingua Franca role to facilitate interpersonal communication among team members (Wang 2015). In such situations, a substantial amount of members have to adapt themselves to speak English, which is a foreign language for them. For example, in OpenAI's **gym** project, at least 20 out of the top 50 contributors (40%) were not native English speakers[1], while its entire repo was purely in English. No matter how good their language skill is, non-native English speakers still often exhibit some noticeable and subtle differences in language use, in contrast to native speakers, particularly in context-specific scenarios (Müller 2005).

Moreover, not all non-native English speakers in a GSD team are good at English. They usually have various levels of English proficiency–the ability to understand and use the linguistic and semantic elements that constitute and transfer meaning in English (Brannen et al. 2017). Some developers could master English as a native speaker, while others had to rely on translation apps. Organizational behavior researchers have revealed that such differences in English proficiency have profound implications for individuals' entire careers in their organizations (Wang and Wang 2011). In general, individuals with high general language proficiency gain certain "language-premiums" such as higher power-related social status (Neeley and Dumas 2016), better interpersonal relationships (Li et al. 2022; Wilmot 2017), etc., in their career.

---

[1] The contributor list of OpenAI's gym: https://github.com/openai/gym/graphs/contributors, accessed on July 20, 2023.

Unfortunately, language issues in GSD teams were largely neglected in CSCW and SE literature, partially resulting from the emphasis on the technical merits in computing professions (Eckhardt et al. 2014; Noble and Roberts 2019). While most extant literature in other fields is based on studies in large multinational corporations (MNCs) (Jonsen et al. 2013), GSD practices happen at more diverse scales in more dynamic organizations. We have little knowledge about whether the results in the literature are still applicable to the GSD. Moreover, an important form of GSD is open source development, which is driven by informal communities rather than traditional organizations (O'Mahony and Ferraro 2007). Open source's ideological underpinnings– advocating equality among members (Singh et al. 2022)–might also offset the power/status imbalances resulting from language capabilities. Its members are usually voluntary contributors rather than contracted employees. Besides, the software development profession's many characteristics, e.g., the extreme imbalance of gender compositions, are naturally woven with language issues, enabling further opportunities for empirical findings, theory development, and practical interventions.

We thus sought to develop an understanding of the experience of GSD developers who were non-native English speakers. To achieve this goal, we interviewed 27 Chinese developers in commercial software development and open-source GSD teams and applied Bourdieu's capital-field-habitus framework in an abductive data analysis process. Our study reveals four types of capital (language, social, symbolic, and economic) involved in their experiences and examines the interrelations among these capitals. We found that those developers' insufficient language capital played an essential role in preventing them from accessing and accumulating other types of capital, thus reproducing the sustained and systematic disadvantaged positions of non-native English speakers in GSD teams. This study contributes to the literature by focusing on a long-neglected aspect of GSD practices and the dynamic interrelations among language, social, symbolic, and economic capitals, explaining why and how language capabilities create and sustain uneven power and status among developers with different native languages in GSD teams. Specifically, our study's contributions to CSCW and Software Engineering (SE) literature are as follows:

1. The study developed an in-depth understanding of the experiences of GSD developers who were non-native English speakers by combining empirical evidence and extant theories. Particularly, our study explains how non-native speakers' insufficient language capital reproduces their sustained and systematic disadvantaged positions in GSD teams.
2. The study introduced Bourdieu's capital-field-habitus framework as the theoretical underpinnings for this study. We demonstrate that applying such a framework would not only benefit the current study's theoretical development but also have the potential in a broad spectrum of research into human and social aspects of various GSD phenomena. Moreover, applying it in GSD contexts, especially open source development, also allows further development of Bourdieu's framework by connecting it with rich CSCW and SE literature, such as the "open fields" identified in our study.
3. The study employed abductive reasoning to analyze qualitative empirical data and integrate it with theoretical knowledge, presenting an example of such a methodological approach and demonstrating its value in qualitative research.
4. The study derived a rich set of theoretical and practical implications for future research and potential interventions for improving practices in GSD teams.

Please note that we use native vs. non-native English speakers, a dichotomous notion in this article. We must admit that such a notion has its inherited limitations because it may oversimplify the diversity within non-native and native English-speaking communities. The language issues are naturally woven with many other social, economic, cultural, and technical factors. Since we focus on a relatively homogeneous sample of participants, we follow the conventions in literature to use such a notion. However, we urge that readers have a critical attitude towards such a notion and its potential limitations. Moreover, the results and findings reported in this article are based on the participants' experiences rather than universally applicable global realities.

The rest of this article proceeds as follows. Section 2 briefly reviews the related work. Section 3 introduces Bourdieu's theories which serves as the theoretical foundation of our work. Section 4 presents the research methods. Section 5 provides a high level preview of the results and findings, with more details in Section 6. Section 7 discusses the implications and limitations. Section 8 concludes the article.

## 2 Related Work

This section briefly reviews the related work of the study. We start with introducing the CSCW and SE literature related to working language in teams. Then, we extend the review to some relevant literature from multiple disciplines.

### 2.1 Studying Working Language in CSCW and SE Literature

The issues around spoken language have been noticed by CSCW researchers, who focus on online communication happening in multiple venues, including social media and work teams. For example, Androutsopoulos (2015) described Facebook users' multilingual practices; Eleta and Golbeck (2012) identified multilingual Twitter users' role in connecting different language communities. Lim and Fussell (2017) examined how multilingual speakers make sense of foreign language posts in social media. Meanwhile, imbalances among speakers of different languages emerged when analyzing multilingual speakers' language practices. For example, Karusala et al. (2018) found multilingual Indian users, though they might not have high English skills, engage in English communication proactively and enthusiastically because using English brought more social and economic benefits to them. To facilitate communication among speakers of different native languages, researchers designed a number of systems and studied such systems' effect on multilingual communication (Lim 2018). For example, Gao and her colleagues (2014, 2015) delivered a series of studies in integrating machine translation into online collaboration, which offered insights and guidelines to the design of such techniques, e.g., providing two machine translation outputs rather than a single output. However, employing machine translation tools is not problem free. For instance, Lim et al. (2022) found that native speakers interpreted the social intentions of the email sender less accurately for machine-translated emails than for emails written by non-native speakers in English and rated senders and emails less positively overall for machine-translated emails. Facing the misunderstandings caused by machine translation techniques, researchers explored user strategies for identifying and recovering from mistranslations (Robertson and D´ıaz 2022).

Meanwhile, SE researchers have long known the issues related to English as the working language in software development, particularly in the contexts of globally distributed development teams. As early as 2000, Bryant (2000) had claimed: "*. . . for good or ill, the language of software development is English–albeit the US variant*" Bryant (2000). Lutz (2009) later introduced the idea of ELF (English as a lingua franca) and presented the high-level challenges associated with it in his conceptual paper. However, it is more or less surprising that such issues were less frequently investigated in SE literature as the major focus. Most extant SE literature treated language issues as a part of other bigger problems, for example, communication or coordination. For example, Holmstrom et al. (2006) pointed out language problem was an important challenge faced by GSD teams in the socio-cultural dimension, and Noll et al. (2011) argued that the linguistic distance is one of the four major barriers in GSD collaboration.

Only a very limited number of studies took the SE team's working language (almost always English) and its influences on the team process and individual experiences as the main research targets. An example of such studies is Wang (2015), in which the author investigated the impacts of the policy of setting English lingua franca on the teamwork in Chinese software outsourcing vendors. The study found that enforcing the English lingua franca may lead to: (1) lower satisfaction at work, (2) a decrease in teamwork quality, and (3) a lack of coordination in terms of socialtechnical congruence. Besides, Bregolin (2022) investigated the effects of allowing the use of multiple languages in programming Q&A communities such as StackOverflow and found that the answer quality and overall participation could be improved.

SE researchers were more interested in designing tools to support automatic multilingual communication translations and facilitate communication among developers speaking different native languages. They often resorted to the same family of techniques as CSCW researchers did, i.e., Machine Translation. For example, Calefato et al. (2016) designed and evaluated a real-time machine translation plug-in tool for text-based online multilingual meetings–eConferenceMT. Their extensive experiments showed that state-of-the-art machine translation technology was already a viable solution for text-based multilingual group communication, although the translations are far from perfect. It was not disruptive of the conversation flow, did not prevent a group from completing complex tasks, and even made discussions more balanced among speakers of different languages. Moreover, their series of explorations of machine translations in SE scenarios provided valuable design guidelines for such systems (Calefato et al. 2010, 2016).

## 2.2 Studying Working Language Beyond CSCW and SE

While the issue of using a specific language in globally distributed teams only receives limited attention in CSCW and SE literature, it has been drawing increasing attention from researchers in multiple other disciplines (Tenzer et al. 2017), particularly scholars in management areas such as organizational behavior and international business. In 2005, while pointing out that studying language issues in the business community was still in its infancy, Ehrenreich (2009) highlighted the importance of exploring how naturally spoken languages affect various processes and outcomes in global organizations epresented by multinational corporations (MNCs). In general, the extant literature is focused on two aspects of languages. The first is about its functional aspect. For example, Fiset et al. (2023) presented a comprehensive study of language-related misunderstanding and its effects on individual and collective performances. The second is about its social aspect. For example, Tenzer et al. (2014) studied how individuals' cognitive and emotional reactions to language barriers influence their perceived trustworthiness and intention to trust in multilingual teams; Neeley and Dumas (2016) linked social status with language barriers and found that native English speakers were likely to gain higher status than their non-native peers in multinational companies.

Recent studies also went deeper into some specific, subtle linguistic features. These features include accents, culture-specific language elements, informal jargon, and so on. Language accents have received particular attention since it is hard for foreigners to excel and are often associated with biases influencing many workplace decisions. For example, Hideg et al. (2022) pointed out the importance of studying the impact of having a non-native accent in the workplace. Moreover, Geiger et al. (2023) designed an experiment and revealed that speaking with a non-native accent (vs. no accent) of English negatively affected hiring decision-makers perceptions of trust and the ability dimension of trustworthiness but not the benevolence and integrity dimensions. In addition, culture-specific language elements also attracted significant attention (Peltokorpi and Xie 2023).

## 2.3 Summary

From the above short literature review, it is fair to conclude that the research into the language issue in GSD teams is limited. At the same time, it has been an emerging topic in disciplines such as organizational behavior and sociology. Current CSCW and SE research on language issues has been fragmented, scattered, and difficult to synthesize. Our study was designed exactly to fulfill this gap. Unlike the prior work in CSCW and SE, we introduce Bourdieu's capital-field-habitus framework as the theoretical foundation of our work, making us reason the language issues more systematically. Thus, our work could offer a unified picture rather than fragile pieces of evidence. Moreover, doing so would enable us to go beyond merely documenting phenomena around spoken language in GSD teams. The theoretical lens leads us to uncover how language capital plays a central role in shaping non-native speakers' experiences. Besides, our work focuses on specific contexts (GSD teams), which have many distinct characteristics compared with related literature in other domains, which usually focus on traditional MNC settings. In particular, open source development has been a major form of GSD. In open source development, cultural, linguistic, and social practices inherently differ from those in traditional MNCs (Han et al. 2024; Von Krogh et al. 2012; Raymond 1999). They are also well beyond the traditional corporation boundaries (Munir 2018; Zobel and Hagedoorn 2020). Moreover, since CSCW and SE have a long tradition of studying open source, our study offers an opportunity for further theory development by connecting Bourdieu's theoretical framework with rich literature and a long tradition of investigating open source in CSCW and SE (Crowston et al. 2008; Ducheneaut 2005; Germonprez et al. 2018).

# 3 Theoretical Foundation: Bourdieu's Capital-Field-Habitus Framework

Non-native English speakers' experiences in GSD teams could be far more complicated than can be determined by their characteristics and actions, for example, their technical knowledge and skills. Researchers have identified substantial influences of the various elements of the context, including many interpersonal, organizational, and social factors, in how non-native English speakers' experiences unfold in international teams (Peltokorpi and Xie 2023; Tenzer et al. 2014). Consider the following case:

> A non-native speaker's English skills are well beyond the requirement of daily software development tasks, but still could not understand some culture-related informal languages, such as some jokes. In some situations, he or she may feel rejected as an outsider of the team.

In this case, developing an in-depth understanding of the speaker's feeling of rejection needs a perspective that explores the interrelationships between himself and the context.

To this end, Pierre Bourdieu, the eminent French sociologist, proposed the capitalfield-habitus framework in one of his classic writings– "*Outline of a Theory of Practice*" (Bourdieu 1977), as a critique to the overemphasizing individual agency[2] in organizations[3]. Therefore, Bourdieu argued that actors (i.e., individuals) cannot be separated from the context where they are embedded in exploring individual-context interrelationships. He thus developed a theoretical framework that emphasizes a relational perspective between actors and their contexts. The framework consists of three interrelated concepts, which are *capital*, *field*, and *habitus*. We are going to explain these three concepts and their relationships before discussing how they could be used to elicit an in-depth understanding of non-native English speakers' experiences in GSD teams.

## 3.1 Capital

In Bourdieu's arguments, capitals are multi-dimensional and no longer only related to economic assets and resources. Capital could be in many types; some frequently used capitals include, e.g., social capital, cultural capital, in addition to economic capital. Social capital was defined as all actual and potential resources that can be accessed through a network of relationships (Bourdieu 1977, 1991). Cultural capital refers to members' various kinds of knowledge, skills, and behavior as members of a specific social group (Bourdieu 2018). Different types of capital are convertible; for example, people's social capital may convert to economic capital if they have some social connections with some senior people who may be willing to promote them. Different types of capital may eventually contribute to enabling individuals to gain certain power in their organizations. Such gained power often takes the form of possession of prestige, status, and a positive reputation, and undoubtedly a type of capital when it comes to the "recognition" of power according to particular existing "schemes." This type of capital was referred to as symbolic capital (Bourdieu and Wacquant 2013). To sum up, capitals are the basis for power in Bourdieusian literature (Bourdieu 1991). Possessions of capitals or a combination of them place individuals in social hierarchies and allows them to influence the valuation of capitals in specific contexts (fields), which further determines shared practices (habitus) of members (Houston 2002).

---

[2] The concept "individual agency" refers to the situation that individuals act at their own will to fulfill their potential and goals.
[3] Bourdieu's theories can be applied to social entities of any scale, e.g., the entire society. However, our study uses "organization" to limit the scope.

## 3.2 Field

Capitals are not necessarily to have definitive or universal values until their realizations in a specific social context. For example, social capital plays a role in people's career development in their organizations (Seibert et al. 2001). However, developers' social capital with their senior managers might have little value in situations such as applying for graduate schools. As an analogy of its use in Physics, the term field was used by Bourdieu to represent the social contexts where capitals derive their value. In fields, individuals strategize and struggle over the unequal distribution of valued capital and/or over the definitions of valued capital by shared rules. These rules are mostly influenced by people having substantial capital and tend to accumulate more capital for themselves (Bourdieu 1991). Therefore, fields are the context for actors to realize and convert their capital, as well as the venues for actors to compete for power. Individuals with limited capital often have to accept the rules, passively take what is left for them, and wait for opportunities (Neveu 2018). A social context may have multiple fields. For example, in the context of an open source project, the legal field (dealing with copyright and license) (Colazo and Fang 2009; Cui et al. 2023) and the professional field (inherited from the commercial companies sponsoring the project) may coexist. A field may also have one or more sub-fields. Each field has its own rules. The rules in different fields may reinforce or contradict each other, jointly affecting the values of individuals' capital.

## 3.3 Habitus

Bourdieu introduced habitus as a set of dispositions to act in a certain way (shared practices), produced by individuals' internalization of social expectations and value systems in relation to the specific field (Bourdieu 1991), i.e., following rules in their actions without a conscious attempt to do so. Habitus enables individuals to display the correct behavior and practices to fit the dominant group's "*taste*." For example, in a GSD team, even developers who were non-native English speakers automatically rather than intentionally use English to communicate with other members. Habitus could be either individual or collective. Individuals may acquire multiple habits as part of groups based on many factors such as gender or social class, and, of course, including their languages.

## 3.4 Relationships among the Three Concepts

Bourdieu's three concepts are interrelated and dynamic. In an organization, dominant groups use different types of capital to construct and maintain social status and power distinctions compared to others, legitimizing and protecting their position in their social hierarchy in the organization. Actors' involvement in fields shapes their habits, which then determine their acts on many occasions. Once the members acted as expected, those fields and the rules of them were reproduced. Through this process, the theory illuminates how the embeddedness of actors in a field makes them legitimize, reproduce, and reinforce the rules by performing socially expected habitus.

## 3.5 Bourdieu's theories, Language, and GSD

As we mentioned above, investigating the experience of non-native English speakers in GSD teams should not separate them from their contexts. Bourdieu's framework thus offers an excellent theoretical foundation for our inquiry because of its unique features, which are summarized as follows.

First, Bourdieu's theories argue that language could be a medium of power by displaying certain competence to facilitate people to compete for power and status through accumulating capital. While GSD teams are the places where languages exhibit strong diversity, they often choose English as their mandatory working language. Such tension naturally produces a dominant group (members who are native English speakers) and a disadvantaged group (members who are non-native English speakers). The dominant group is in a privileged position to gain multiple types of capital and realize their values. Therefore, the theories empower us to go further than describing the phenomena around non-native English speakers in GSD teams by helping us formulate plausible explanations for these phenomena under an abductive logic.

Second, Bourdieu's theories are not only theories but also thinking tools (Costa and Murohy 2015). The key theoretical constructs in Bourdieu's theories, namely, capital, field, and habitus, offer a critical perspective to guide our empirical inquiry and bridge the theory and practice. Let us take the concept habitus as an example. Together with the empirical evidence, Bourdieu's framework allows us to explain how and why developers who are non-native English speakers conceive and (re)construct the social realities in GSD teams. Meanwhile, it also directly links to those non-native speakers' internalized behaviors, perceptions, and beliefs that are translated into their practices in GSD teams. Such connections between theories and practices would help us formulate practical implications, e.g., which habitus shall be encouraged to address issues related to language use?

Lastly, Bourdieu's theories are naturally dynamic, open, and relational. Its dynamic feature allows the description of a complex social process in which developers and their contexts together construct dispositions to justify ever-changing perspectives, values, actions, and social positions while also capturing, reasoning, and theorizing the invariant features.

To sum up, Bourdieu's theories provide an excellent theoretical foundation for our study of non-native speakers' experiences in GSD teams.

## 4 Research Methods

### 4.1 Research Design and Contexts

Since the theme of this research remained relatively unexplored, a qualitative study would help us develop a deep understanding first. It also better served our purpose to learn complex dynamics and rich contexts around the non-native English speakers' experiences in global software engineering. We conducted the study with software developers from China. We chose China for several considerations. First, belonging to the Sino-Tibetan languages, Chinese has a long linguistic distance away from English, an Indo-European language (Thurgood and LaPolla 2016). Thus, it is difficult for Chinese to learn and excel in English. Second, Chinese is a high-context language while English is a low-context language (Duranti and Goodwin 1992). Third, many international software development organizations have research and development operations in China. Though not always setting English as a lingua franca, these organizations have many opportunities for local Chinese software developers to engage in global development teams. Finally, China has one of the largest populations of open source contributors. For example, only in 2022, there were over 1.2 million developers from China joined GitHub[4]. These characteristics made China an ideal location for us to study GSD teams' non-native English speakers' experiences.

## 4.2 Participants and Data Collection

We collected data through semi-structured interviews. This section briefly introduced how we recruited participants and performed the interviews.

### 4.2.1 Participants

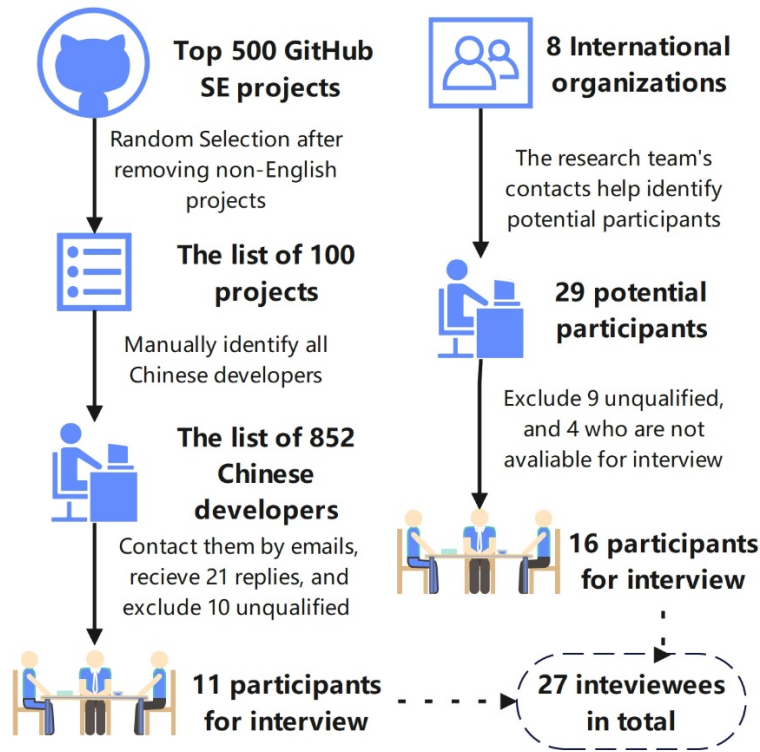

Fig. 1 The process of recruiting participants from both open source projects and commercial organizations.

We used a purposeful sampling approach (see Fig. 1) to recruit interview participants satisfying the following four criteria: (1) they must be native Chinese speakers and have never spent more than a year in English-speaking countries; (2) they must spend a substantial amount of time (≥ 20 hours per week) working in GSD teams, no matter whether the time was spent in working for commercial software development organizations or contributing to open source projects; (3) English must be the working language in their GSD teams; and (4) they must belong to minorities regarding the native language, i.e., their GSD teams should have more native English speakers than non-native English speakers.

---

[4] https://octoverse.github.com/2022/global-tech-talent

**Table 1** Basic information of the participants.

| No. | Org.* | Role** | Age | Exp.@Org. | Exp.@GSD | Sex |
|---|---|---|---|---|---|---|
| P1  | C | N   | 20-30 | 0-5   | 1-5   | M |
| P2  | C | N   | 40-50 | 0-5   | >10   | M |
| P3  | C | M/L | 30-40 | 5-10  | 1-5   | F |
| P4  | C | M/L | 30-40 | 0-5   | 1-5   | M |
| P5  | C | N   | 30-40 | 10-20 | >10   | M |
| P6  | C | N   | 20-30 | 0-5   | 1-5   | M |
| P7  | C | N   | 30-40 | 10-20 | >10   | M |
| P8  | C | N   | 20-30 | 0-5   | 1-5   | F |
| P9  | C | N   | 20-30 | 5-10  | 5-10  | M |
| P10 | C | N   | 30-40 | 0-5   | >10   | M |
| P11 | C | N   | 30-40 | 5-10  | 5-10  | M |
| P12 | C | N   | 20-30 | 0-5   | 5-10  | M |
| P13 | C | M/L | 20-30 | 0-5   | 5-10  | M |
| P14 | C | N   | 30-40 | 5-10  | >10   | M |
| P15 | C | N   | 20-30 | 5-10  | 1-5   | F |
| P16 | C | N   | 30-40 | 5-10  | 5-10  | M |
| P17 | O | M/L | 30-40 | 5-10  | >10   | M |
| P18 | O | N   | 20-30 | 0-5   | 1-5   | M |
| P19 | O | N   | 20-30 | 0-5   | 0-5   | M |
| P20 | O | N   | 20-30 | 0-5   | 0-5   | F |
| P21 | O | N   | 20-30 | 0-5   | 5-10  | M |
| P22 | O | M/L | 20-30 | 0-5   | 5-10  | M |
| P23 | O | N   | 30-40 | 0-5   | >10   | M |
| P24 | O | M/L | 20-30 | 0-5   | 0-5   | F |
| P25 | O | M/L | 20-30 | 5-10  | 5-10  | M |
| P26 | O | N   | 30-40 | 0-5   | 5-10  | M |
| P27 | O | N   | 20-30 | 0-5   | 5-10  | M |

*. C: Commercial organizations, O: Open source projects.
**. M/L: Taking project management/leader role in current GSD team, N: Not.

The recruiting of participants consisted of two parallel processes for it is crucial for this type of study to involve samples of developers from GSD teams in both commercial and open source software development organizations. The first was used to recruit participants whose GSD experiences were mostly in commercial software development. We first identified a few multinational software development organizations and accessed potential participants through the research team's personal contacts. The second process was used to find open source developers who satisfy the above criteria. We manually identified the Chinese developers from the 20 projects randomly selected from the top 200 software development projects on GitHub according to their numbers of stars. Then, we asked for their basic information related to the above criteria and their willingness to participate in the study via email. Eventually, we had 16 participants (P1-P16) from commercial software development organizations and 11 participants (P17-P27) from open source projects. Tab. 1 listed their basic demographics.

In addition to the demographic information in Table 1, all 27 participants had at least 4-year college education, mostly in STEM majors. Among them, 16 received master-level postgraduate or above education. All passed the CET-4[5] English proficiency test, while 22 passed the CET-6. Therefore, it is fair to say that our participants had at least intermediate-level English skills, which allowed them to have the necessary fluency to communicate without effort with native speakers. Note that the standard test scores are only an imperfect reflection of one's real English capability. It is not uncommon for people having similar scores to exhibit quite different levels of English fluency.

### 4.2.2 Interviews

We conducted seven face-to-face and 20 online video/audio interviews. Among the 27 participants, 23 consented to allow us to record their interviews and keep the digital recordings for an extended period of time. The interviewers took detailed notes for the other four interviews. We began interviews with several questions about the participants' demographic background and the organizations where they participate in global software engineering. We continued with open-ended questions related to their work experiences. A few examples of these questions include: (1) What were the reasons for you to work in GSD teams? (2) What are your experiences working in GSD teams? (3) What are the challenges when working in GSD teams? We asked follow-up questions and encouraged the participants to elaborate on their narratives. We let the topics related to experiences as non-native English speakers emerge naturally along the process of interviews rather than directly asking specific questions before the participants mentioned them. We conducted interviews in Mandarin Chinese. The recorded interviews were transcribed, forming the primary documents for data analysis with the interview notes. Since the data would be analyzed by the authors who are native Chinese speakers, we did not translate the primary documents into English. Instead, we selectively translated some pieces of the participants' narratives when drafting this manuscript.

### 4.5 Data Analysis

Different from the common inductive logic in analyzing qualitative data, we followed abductive reasoning in our data analysis. Abductive reasoning (Fig. 2.c) is defined as "a syllogism in which the major premise is evident but the minor premise and therefore the conclusion only probable"

---

[5] CET contains two standardized English proficiency tests at different levels (CET-4 and CET-6) for Chinese college students not majoring in English. Passing CET-4 is roughly equivalent to an IELTS score of 5.5 or CEFR B1 (Kang 2022).

(Flórez 2014). It lies in the middle between deductive reasoning and inductive reasoning, but has been largely neglected by scholars. It does not aim at developing general conclusions by analyzing empirical observations of singular events, which inductive reasoning (Fig. 2.a) usually does (Richardson and Kramer 2006). Different from deductive reasoning (Fig. 2.b), it does not first select a theory as a general rule and then draw specific conclusions through the formulation and testing of hypotheses (Haig 2005). In abductive reasoning, the goals are to generate probable hypotheses that indicate incremental knowledge development by combining theoretical and empirical evidence (Dubios and Gadde 2002). Therefore, it fits our research goal well. However, readers should keep in mind that situating abductive reasoning in qualitative data analysis has its fallible nature, as other types of reasoning (Lipscomb 2012).

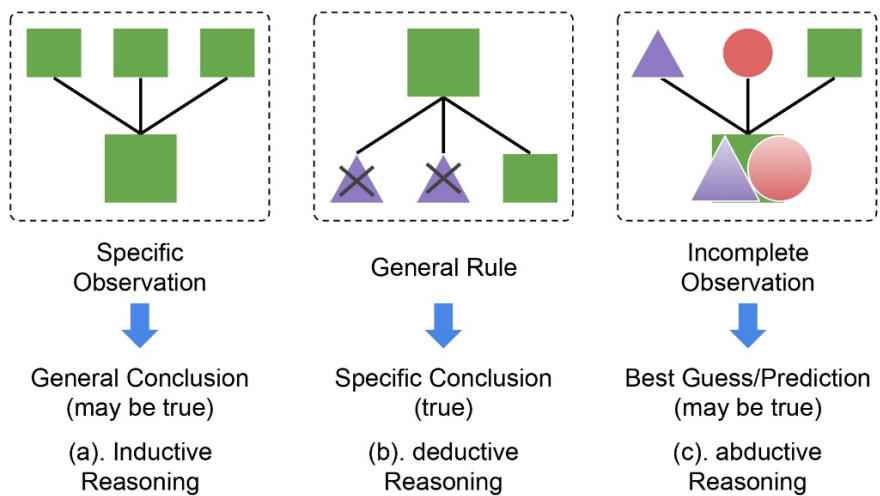

**Fig. 2** Simple illustrations of three types of reasoning.

The findings of the study were produced based on Bourdieu's capital-field-habitus framework, alongside an increasingly detailed analysis of the interview data through several coding phases adapted from the grounded theory (Sætre and Van de Ven 2021; Tavory and Timmermans 2019; Timmermans and Tavory 2012). Since we took an indirect, natural way to lead participants to tell us their experiences in GSD teams related to language, a substantial proportion of our data was not quite relevant to the focus of the study, for example, their narratives about the technical details about projects they are working on. Therefore, we first read through all interview recordings and notes and coded them into one super category that covered all language-related content. Such a coding procedure was performed in the presence of all authors. We took an inclusive strategy in this process, i.e., if a piece of data was assumed to be relevant by anyone in the author team, it would be included in the super category.

Then, the data was analyzed through a standard qualitative data analysis procedure. Two authors of native Chinese speakers coded the filtered data independently through an *Open Coding* process when provisional codes were assigned to them based on the meanings identified by the coders. During this process, both coders met together almost every day to discuss the code and resolve the conflicts collectively to ensure the correctness of the codes in interpreting participants' meanings. In some very rare cases of lack of agreement on certain codes, the third coder, who was also a native Chinese speaker, was invited to make the arbitration. Doing so helped expose each coder's interpretive mistakes as early as possible, thus avoiding potential misunderstandings and inconsistencies resulting from coders' different understandings to be real, long-lasting interpretative threats (Evans 2013). In *Axial Coding*, the identified codes were put together and examined again to find the connections among them, which helped us cluster some concepts together as categories. Then, we further identified the properties and dimensions of these categories. Later, we returned to the data and re-coded them with these concepts. *Constant comparisons* were performed from the beginning and throughout the entire process of data analysis. I.e., the emerging concepts were repeatedly compared with old and new data. The *Selective Coding* focuses on critical categories and concepts. The selection was based on the frequency of codes' occurrences or the patterns they appeared in the data, e.g., some were frequently mentioned by multiple participants.

The final step was *Theoretical Memo Writing and Concept Refinement*, in which the theoretical knowledge was developed with the space created through theoretical memo writing, we compared the data, codes, categories, and concepts in our analysis and focused on the theoretical propositions linking them together (Muller and Kogan 2012). Then, we refined the emerging categories and concepts. The analysis found repeated appearances of four themes around the experiences of GSD developers who were non-native English speakers: language capabilities, social interactions, status and power, and individual benefits such as career advancement. Then, guided by the capital-field-habitus framework in Bourdieu's theory of practice (Bourdieu 1977), we coded these topics as (1) language capital, (2) social capital, (4) symbolic capital, and (4) economic capital. Each contains a few data categories, which will be introduced in the findings in detail. Moreover, our analysis also discovered the interrelations among them and formed explanations for the role language played in continuously placing non-native speakers in disadvantaged positions.

## 5 Preview of Main Findings

Before presenting the detailed results and findings in the following sections, we provide a preview of them in this section, which is represented by a framework summarizing different types of capitals and their interrelations involved in the non-native English speakers' experiences in GSD teams. Fig. 3 depicts the overall framework.

There are four types of capitals (language, social, symbolic, and economic) in the middle; each is associated with a few sub-dimensions identified from the data. First, the first three could convert to each other. For example, high language proficiency could help one fit into social networks, thus realizing the conversion from social capital to social capital. Meanwhile, being an insider of native speakers' social networks would bring more opportunities to practice English, thus facilitating the conversion from social capital to social capital. Other cases are similar. Second, all the first three could convert to economic capital, but not vice versa. Since the first three capitals are interconvertible, all could convert to economic capital either directly or indirectly with the others as the mediums. In this process, language habitus, as coded behavioral act rules, ensures the social orders defined by the dominant group (native English speakers in GSD teams) are reproduced and reinforced. Together, non-native speakers were put into disadvantaged positions in their organizational fields and open fields (via open source).

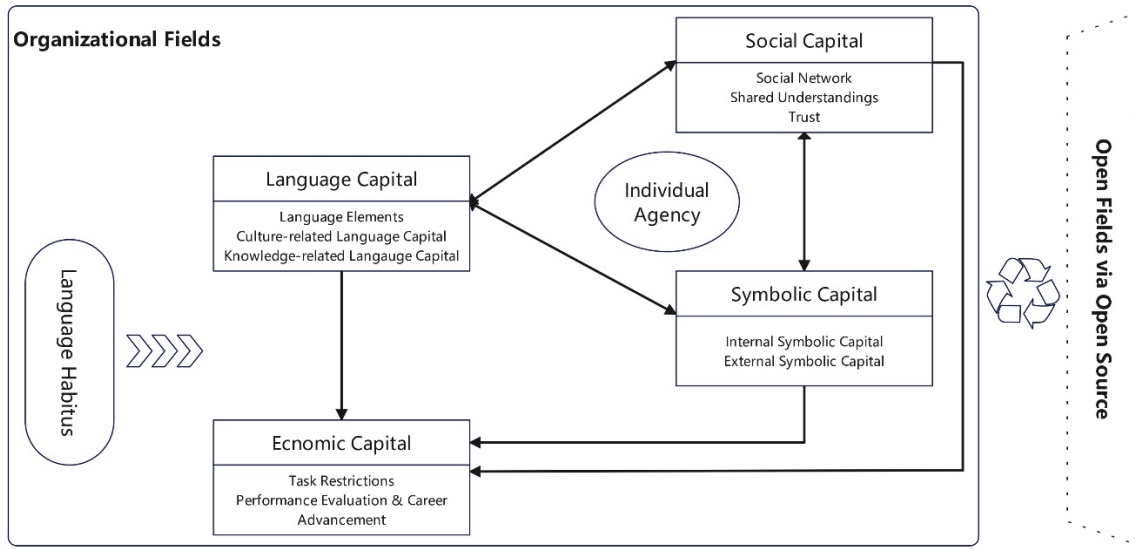

**Fig. 3** The overall framework of findings.

## 6 Results and Findings

We report the detailed results and findings in this section. We will introduce the four capitals and their interrelations together rather than put them in a separate section because they are easily discussed and understood within the context.

### 6.1 Language Capital

In GSD teams where non-native speakers are minorities, developers' language capabilities become the fundamental capital. Insufficient language capital means that non-native speakers have less capital for conversions to other types of capital, which further limits their opportunities to access and accumulate language capital. For native speakers, language capital could be used to differentiate non-native speakers' lower status as outsiders of their circles and to associate language capital with lower job capabilities. Doing so could reproduce and reinforce the asymmetries of power and status between them. Our analysis identified three types of language capital, which are (1) *language elements*, (2) *culture-related language capital*, and (3) *knowledge-related language capital*.

### 6.1.1 Language Elements

A modern language contains many language elements. Among these language elements, two were particularly important in the GSD context. The first is the accent, and the second is the choice of words. The accent was often the most difficult barrier for non-native English speakers and thus could be a significant indicator of one's status. "Native speakers believe accent is your own problem–you did not learn hard or want to learn English," **P11** said. Native speakers reacted even to small differences in accents, making non-native speakers frustrated. The choice of words had similar consequences but was often more serious because text-based interactions in GSD teams were much more frequent. Word choices also determined the tones of people's speech. Non-native speakers tended to use less definitive tones, e.g., saying "maybe" a lot. Such word choices have been proven to be associated with lower social status in software development teams (Han et al. 2023). By emphasizing these language elements, non-native speakers' language capital was discounted, thus reducing the other capital it could convert to.

### 6.1.2 Culture-related Language Capital

Another type of language capital is related to certain cultural practices. Fortunately, as a low-context language, English does not have substantial information embedded in context. However, non-native speakers often use their experiences with their native language in using English. Modern Chinese have already gotten rid of many honorifics that encode the relative social status between speakers. However, it is widely viewed as an impolite act to call one first name if the speaker has lower or equal social status. **P19** said he was shocked when he noticed that project members from the US call each other's first name or screen name directly. He tried to be polite and put honorifics such as Mr./Ms. in front of other members' screen names on their GitHub profiles. Of course, native speakers realized he was a non-native speaker. Some members even mimicked him, which made him feel hurt. Non-native speakers were able to amuse native peers by making fun of their foreignness and inability to fully understand cultural practice reflected in English and its use of language. Therefore, non-native speakers often had to accept their disadvantaged positions in GSD teams. Moreover, while modern language technologies are quite sophisticated, they provide little help in gaining culture-related language capital. Our participants echoed that language technologies such as translation apps in their phones had very limited utility in interpersonal communication with native speakers, "it helps you understand road signs when you are traveling, but does not make your work emails convey the right meanings," **P6** commented.

### 6.1.3 Knowledge-related Language Capital

The last type of language capital is knowledge-related language capital. Note that here, the word "knowledge" does not mean language knowledge but the everyday common knowledge a native speaker could have. When being touched in conversations, this knowledge would immediately signify the differences between native and non-native speakers. Native speakers do not intentionally learn such knowledge but get it in their everyday life. For instance, it is easy for Americans to name some NFL teams even if they are not football fans. However, non-native speakers have to make extra effort to learn such knowledge. Otherwise, they would not be able to join native speakers' conversations. **P5** told us:

> You know Super Bowl, American's party festival, like Chinese Spring Festival Gala. Messages about it were flooding our Slack channels for weeks. I had to spend several hours on it to say something, to make them believe I am living in the same world with them.

Owning similar levels of knowledge-related language capital is impossible for nonnative speakers. Such knowledge has overwhelming breadth for non-native speakers who often only obtain technical skills, thus, making it a valuable capital for native speakers that can convert to other capitals, for example, trust in social capital.

## 6.2 Social Capital

The lack of language capital often makes non-native English speakers fail to join the circle of native speakers in GSD teams. Therefore, they experienced difficulties in developing meaningful friendship ties and accessing critical social support from social networks. In addition, the limited language capital also prevented developers who are non-native speakers from developing and maintaining trusting relations with other members of the GSD team.

Our participants reported their failures in obtaining three types of social capital. These are *social network*, *shared understandings*, and *trust*, corresponding to the structural, cognitive, and relational dimensions of social capital, respectively (Andrews 2010).

### 6.2.1 Social Network

Non-native English speakers' insufficient language capital exposes their non-native status and makes them excluded from native speakers' social networks. Developers who are non-native English speakers often feel native speakers treat them as temporary workers with lower status rather than valid members. Such segregation did not always result from the spatial and temporal distances in GSD teams. **P7** once visited his company's headquarters in Silicon Valley for two months before the COVID-19 pandemic; he recalled his experiences there:

> They had a welcome party for me. Unfortunately, I immediately noticed that they socialize selectively. I had some accents and could not pronounce some words correctly. I mispronounced the word "vacation" as "vocation," they noticed it and made some fun of it. They made me feel bad. They came to talk with me, but I felt they just pretended to be welcome and polite. The conversations were very short and quite superficial.

Obviously, even in the same place, **P7** was also isolated by his native colleagues. Some participants felt that was quite understandable. **P3**, a senior manager, shared her observations with us.

> If you ask me if I regret not going to Bellevue six years ago. I would say yes, not for myself but for my kid because she might have a happier life there. I visited there many times. My Chinese colleagues there [Chinese immigrants] were always treated as outsiders. Native people seldom invited them to socialize together. You had to be in the Chinese circle because your English is imperfect. It is quite understandable. We also don't invite foreigners to join our gossip party here [in Beijing]. When you say some fun punchlines that only Chinese can get, they are real killjoys.

An interesting phenomenon is that non-natives also had their own circles where they shared their feelings and strategies for dealing with foreign people. This was an evident sign of individual agency to change their disadvantaged situations as linguistic minorities. "We all had Slack and Zoom meeting, but only we three Chinese had WeChat groups," commented by **P9**.

### 6.2.2 Shared Understandings

Shared understanding among team members is critical for both team and individual success,

particularly for heterogeneous GSD teams (Bittner and Leimeister 2014). Non-native speakers' insufficient language capital often prevented them from developing shared understandings. First, they might not fully understand the situation due to limited language capabilities. For example, **P27** told us that he was "totally lost when reading the project's long documentations describing the workflow." and he suspected that these documentations "were purposefully written in this way to discourage nonnative people." Second, they might interpret specific situations in a way different from what native speakers did, which was often related to culture-related language capital. For instance, multiple participants reported that they experienced a hard time in seeking to understand the native speakers' intention behind their words. Later, they found that most of these difficulties resulted from the influence of their home culture. For example, **P15** told us her story:

> I used to be overly sensitive. As a young woman just graduating from university then, I had zero experience working with an international team. I usually spent an hour figuring out if my manager [in the US] was not happy with me. I often felt very puzzled.

However, shared understandings were defined by the dominant group–native speakers. The non-native speakers were expected to follow them (*habitus*). Failure to develop shared understandings, therefore, made non-native speakers less capable in competing capitals. Thus, the existing social orders could be well-maintained.

### 6.2.3 Trust

While acknowledging non-native speakers' technical capabilities, native speakers in a GSD team often did not trust that their non-native colleagues could fully understand their communication and get tasks done accordingly, partly due to the lack of linguistic capital. A majority of our participants experienced some doubts about their abilities to understand assigned tasks in English from their colleagues. For example, **P15** provided her experiences.

> My former manager usually emailed me after discussing what needed to be done in our scrum meeting. I used to think that he did it to everyone, but it turned out that I was the only one receiving such emails. I think he didn't trust me, either because I was female or I did not speak English well enough.

Our participants also told us that their supervisors often asked another colleague to perform some additional checks on their work, which made them quite uncomfortable. Also, they noticed native speakers often avoided interacting with them. **P10** said, "I was the person receiving the least number of emails in the team, even for autogenerated code review request emails...I can talk to my US colleagues about anything without problems, but they can't."

Obviously, the interaction quality, e.g., frequency, breadth, and depth, between non-native and native speakers in GSD teams is often lower. Thus, non-native speakers usually do not have a chance to develop meaningful, trusting relationships. SE literature has repeatedly confirmed that non-work-related, informal conversations played an important role in multiple stages of trust development (Zheng et al. 2002; Wang and Redmiles 2016). Culture-related and knowledge-related language capitals, rather than language elements, were the main barriers for non-native speakers to reduce the subjective distance from native speakers. **P22**, who was a Project Management Committee (PMC) member in a large open-source project, commented:

> I found a website–Urban Dictionary–very helpful to non-native, non-US people like me. My fellow colleagues from the US used many words in it. It helped me to correctly understand their meanings in jokes, small talk, etc. You see, sometimes you need to know a little bit more. For example, you may need to know something about the English monarchy and the Royal family's gossip to join your colleague's conversations about Charles III's coronation.

Other participants echoed **P22**'s comments. **P12** recalled visiting HSBC's call center in Guangdong, which provided services for Hong Kong and Macao customers. He said, "The banking industry understands this better than us. HSBC's call center is equipped with dozens of large TVs broadcasting horse racing in Hong Kong Jockey Club. So, their staff could talk with customers in Hong Kong as they were there and win their trust." Meanwhile, lacking meaningful conversations not only hindered the development of trustful relationships with native speakers but also restricted their opportunities to accumulate language capital through interacting with native speakers, which made situations even worse. "Anyway, you can't learn English if they don't want to talk with you," **P14** commented.

Moreover, our participants also worried that using modern language technologies would threaten native speakers' trust in them. **P13** told us that he used to rely on Grammarly[6] to fix the language issues in writing some important messages. However, he admitted that he "often feels less confident after using it because he had no idea if the fixes made by Grammarly distorted his initial intentions and meanings." The situation could be even worse. If foreigners' English in formal writings were much better than their English in daily communication, "they will know you using some tools in writing," **P13** added, "they may become judgmental and think you are a disingenuous person." Therefore, the damage to the trust occurred.

## 6.3 Symbolic Capital

As the capital of "honorability," symbolic capital signifies obtaining certain prestige, status, and ranking in a GSD team. Symbolic capital could be *internal symbolic capital* and *external symbolic capital*. Furthermore, the *value realization of symbolic capital in open source* has some unique characteristics.

### 6.3.1 Internal Symbolic Capital

Recognition that GSD developers receive in their organization could be their internal symbolic capital. For example, GSD developers could receive recognition such as "knowledge sharing" awards or "code review" masters. These symbolic capitals are often displayed on their internal profile pages to signify their distinctions. Such recognition establishes power and status in the organization. For example, code review masters often have more opportunities to judge others' work, thus guaranteeing their certain power over others. For non-native speakers, their insufficient language capital resulted in serious barriers to obtaining such capital. Take the knowledge sharing award as an example. In P8's organization, such awards were usually given to people who were active in their internal social network to write technical blogs, sharing viewpoints, etc. P8, though quite confident about her technical knowledge, was not confident enough to write long technical articles in English. P8 told us, "It took me a month to write my first blog. Then I read it; I felt it was not good, boring, honestly. I realized I could not compete with native speakers in writing. That was my only try." Some symbolic capital may be obtained based on certain social capitals, for example, being a project's technical committee member, which further restricted nonnative speakers' chances. Therefore, native speakers' language capital could help them monopolize internal capital directly or indirectly, thus reproducing and reinforcing their advantages in GSD teams.

---

[6] Grammarly is an AI-based cloud writing assistant reviewing language issues and providing suggestions for fixing such issues.

### 6.3.2 External Symbolic Capital

GSD team members may also obtain external symbolic capital. For example, if members were designated as senior-ranking members in prestigious professional societies or had won some prizes, they might also gain symbolic capital. There are two key challenges for non-native speakers. First, the symbolic capital accumulated in their home countries is hard to translate to those in English-speaking countries. Second, the lack of language capital hinders their opportunity to be recognized in International venues. These two challenges often come together and further limit non-native speakers' access to and accumulation of symbolic capital. As **P17** commented:

> I was a distinguished speaker and won many national prizes for my contributions to open source. But these don't make too much sense for my global teammates. Nobody cares. They still congratulate you, but that's it. But getting International recognition is hard. I failed in applying for the IEEE senior member last year, partially because I cannot write English papers or give English talks. Honestly, I had to ask a friend to help me finish my application form.

### 6.3.3 Symbolic Capital in Open Source

Moreover, symbolic capital could have far more profound impacts in open-source development. The "*open*" nature exposes the symbolic capital to the public domain, giving its members opportunities to realize the value of their projects. For example, if they become members of prestigious projects' project management committees, they would have the reputation of being top experts in the relevant technical area. These reputations can easily convert to significant economic capital beyond the project. **P25** told us that he earned more than 60,000 Chinese Yuan [about 8,390 USD] by giving talks and technical briefings on various occasions every month after being a famous data management project's PMC member. This suggests that the conventional meanings of field in Bourdieusian theories must be updated to incorporate such open fields, in which people possess social positions and accumulate certain capitals beyond the traditional boundaries of formal organizations. Such fields also provide new venues for interactions and conversions among different forms of capital.

## 6.4 Economic Capital

The monolingualism of English's de facto priority as a working language enables privilege and dominance of native speakers over non-native speakers. The lack of language capital restricts non-native English speakers' opportunity to gain certain economic capital. Thus, language capital, along with social and symbolic capitals, directly influenced, reproduced, and maintained the economic capital distributions favoring native English speakers in GSD teams. We found there are two interrelated restrictions, which are: *task restrictions*, and *performance evaluation and career advancement*.

### 6.4.1 Task Restrictions

Due to insufficient language capital, non-native English speakers were often locked into a narrow set of tasks. These tasks were often independent, atomic, and trivial ones, requiring minimal interactions with other parties. As we mentioned in Section 6.1.1, non-native speakers were very sensitive to the choice of expressions and words in social interactions, and they often spoke in an indefinite tone. On many occasions, they kept silent if they were unsure what to say. As a disadvantaged group, these had already been their habiti. But, such acts often make their native peers consider them to be less capable and confident. Therefore, they often had little chance to perform challenging tasks. Besides, while contributions to open-source projects are open to any potential contributors, task restrictions were implemented through the review mechanism after task completion.

> I often had some pull requests rejected for typos or grammar mistakes in commit messages or code comments. I understood their intentions to maintain high-quality documentation. But I was frustrated to be rejected for a few misuses of articles, especially when the code they accepted was not as good as mine but in better English.

Moreover, though frustrated, **P26** continued his contribution to the project. The project was very high-profile in the area of data science. If listed as a contributor, a developer may receive significant benefits such as recognition as an expert by potential employers. "You dislike that, but have to accept it," **P26** commented. Thus, the gain of economic capital makes open source's voluntary contribution no longer work. Apparently, "accept it" becomes a habitus for **P26**.

### 6.4.2 Performance Evaluation and Career Advancement

The performance evaluation played a critical role in realizing the value of other types of capital. In GSD teams, insufficient language capital did not directly influence the results of performance evaluations but also had an indirect effect through social and symbolic capital. For example, non-native speakers with limited language capital may not be able to develop trust (social capital) with key people in the team, which would lower their performance evaluation. Meanwhile, they also developed certain habitus in performance evaluation, i.e., accepting whatever they get. As **P12** said, "I don't expect to get a very good evaluation. Only highly likable insiders receive them. I would be fine if not getting a PIP [Performance Improvement Plan]."

Performance evaluations have direct effects on one's financial benefits. They would receive less bonus or fail to get a pay raise if they did not get good evaluations. Moreover, performance evaluations also determine one's career advancement. For example, **P3**'s organization required those who wanted to be promoted to receive at least two "As" in the last three performance evaluation cycles. To sum up, due to limited language capital and scarce opportunities to improve their language capabilities under the native speakers' dominated situations, non-native speakers faced challenges in (re)producing essential language, social, and symbolic capital for the conversions to the economic capital. Therefore, language capital serves as a filter for reinforcing the social hierarchy in GSD teams.

Unlike those working in commercial GSD organizations, open source GSD developers' performance evaluation did not directly link to monetary benefits. However, it significantly influenced their career advancements in a much more open field. In addition to members of a GSD team, external people also had the privilege of evaluating an open-source developer's performance. Rich language, social, and symbolic capital gained from their contributions to open-source projects would help their career advancements in multiple ways. We have mentioned that **P25** received extra incomes with the symbolic capital accumulated in an open-source GSD project team. These capitals would also help attract potential employers (Marlow et al. 2013). However, non-native speakers lacking these types of capital would likely be locked in their disadvantaged positions.

# 7 Discussion

The study reported in this paper confirms many findings in the existing literature. First of all, it provides concrete and convincing evidence to show that language issues are still one of the major challenges in GSD practices (Holmstrom et al. 2006; Lutz 2009; Noll et al. 2011), after several decades of GSD practices. Second, it connects findings dispensed in rich multidisciplinary literature in a unified framework and relates them to GSD practices, e.g., individual motivation vs. individual agency (Karusala et al. 2018), language elements (Hideg et al. 2022), economic capital in open source (Marlow et al. 2013) and so on. Meanwhile, our study brings many new findings, e.g., the problem of translating home-country symbolic capitals to English-speaking countries, the link between trust and using modern language tools, etc. Moreover, the integration with Bourdieu's theories enables us to go deeper than only describing phenomena related to language in GSD teams. It allows us to reveal that non-native speakers' insufficient language capital played an essential role in prohibiting them from accessing and accumulating other capitals, thus reproducing the sustained and systematic disadvantaged positions of non-native English speakers in GSD teams. In this section, we first briefly discuss the study's theoretical, methodological, and practical implications. Then, we discuss the threats to validity.

## 7.1 Theoretical Implications

We now briefly discuss the theoretical implications of the study. We focus on three critical points: (1) the introduction of Bourdieu's theories, (2) the use of abductive reasoning, and (3) language diversity's position in the emerging area of software development diversity and inclusion.

### 7.1.1 Bourdieu's Perspective as Theoretical Underpinnings for Researching Global Software Engineering

First, this study introduces Bourdieu's capital-field-habitus framework as the theoretical lens to analyze spoken language issues in GSD practices. It fruitfully employed Bourdieusian analysis to elucidate how GSD developers' non-native English proficiency could be barriers to accessing and accumulating career-critical capital, which in turn reproduced and reinforced their disadvantaged positions in GSD teams. Therefore, we contribute significantly to the GSD literature by offering such a theory-grounded understanding with empirical support. Our study also contributes to Bourdieu's theories by extending its applications to GSD domains with novel extensions. Bourdieu's theory does not view individual agency as a strong force in determining an individual's social situations, and gives it limited space in its theoretical construction (Bohman 1997; Elder-Vass 2007). While our study confirmed that being non-native English speakers limited their capability in accessing and obtaining capital, some participants still try their best to improve their situations (see **P22**'s case). Such an individual agency, which coincides with the findings in Karusala et al. (2018), could be connected to a rich body of individual motivation literature in CSCW and SE.

Moreover, we found symbolic capital could be realized outside of the organization for open source projects, which is rarely reported in Bourdieusian literature but well-known in CSCW literature, though have not yet been conceptualized as a capital (Bosu et al. 2014; Dabbish et al. 2012). This indicates an opportunity for both Bourdieusian literature (mostly from the sociology domain) and CSCW literature to learn from each other to develop an in-depth understanding of specific phenomena and build theories. First, this implies that Bourdieusian theory could be enhanced by incorporating the concept of "open fields." This conceptualization has particularly theoretical importance since open innovation and social production, represented by open source development, are steadily rising in the modern organization of production (West and Bogers 2014). Open fields provide venues for the interactions among language capitals and other capitals, such as the external symbolic capitals, which may create new power dynamics and social hierarchies. As we mentioned in Section 6.3.3, certain capitals in open source can compensate language capitals and be directly realized to economic capitals. Therefore, our work modernizes Bourdieusian theories to some degree.

Second, Bourdieu's theories could be applied to many other topics in human and social aspects of software engineering. The framework and the related theoretical literature are general-purpose social theories concerning the interrelationships between the actor and the contexts (Hilgers and Mangez 2014) and have been successfully applied in many disciplines. Therefore, they could be applied to critically analyze any CSCW and SE phenomena and practices involving various stakeholders (i.e., actors) and rich contexts such as governing structure, team composition, etc. Our work serves as an example of mounting empirical evidence from the Bourdieusian perspective to develop a deep understanding of issues in globally distributed development. There are many other potential areas in CSCW and SE where Bourdieusian analysis could contribute theoretically grounded novel explanations. For instance, consider the emerging area of gender diversity in software development. While the extant work has accumulated rich empirical evidence, Bourdieusian analysis could inform researchers about theoretical understandings of women's disadvantaged positions in capital access and accumulation by reexamining the empirical evidence and identifying actionable remedies.

### 7.1.2 Abduction as a Novel Methodological Approach in CSCW

We employed abductive reasoning rather than inductive reasoning–the common methodological approach for qualitative data analysis–in data analysis. While some literature has connected abductive reasoning with grounded theory or other qualitative inquiries (Tavory and Timmermans 2019), such an epistemic process is rarely used by CSCW researchers. The abductive approach allows for the inclusion of theoretical knowledge, providing guidance while simultaneously investigating participants' narratives and extending understanding beyond known effective approaches. In our study, it at least offered several benefits. First, it helped generate deep insights from interview data by analyzing and mapping data to the preexisting Bourdieusian literature. Without such a mapping, it would be hard to link our participants' narratives with theoretical constructs such as *capital*, *field*, and *habitus*. Second, it provided a framework for incremental theory building, i.e., refinement of existing theories rather than building new ones from scratch. Through a cross-fertilization process, novel enhancements were developed through a mixture of existing theories and new data. For example, as mentioned above (see Section 7.2.1), data from open source GSD team members indicated the necessity of extending the conceptualization of the field in Bourdieu's framework. Lastly, it enabled us to develop rich types of theories since abductive reasoning is also capable of generating theories of explanations and predictions, or even prescriptions (Gregor 2006).

### 7.1.3 Diversity and Inclusion in Software Development

Finally, our work belongs to a growing body of literature on diversity and inclusion in software development. The diversity issues around spoken languages receive much less attention than frequently discussed gender or race issues. We argue that it is imperative for CSCW and SE research to put more effort into investigating language diversity. First, more and more developers from non-native English-speaking countries are joining the global software engineering workforce. Addressing language barriers would not only benefit these new populations of software development workforce but also attract and retain global talents for software development organizations. Second, language is not an isolated construct but a medium connecting many constructs such as human cognition, country of origin, economic background, culture, and behavioral decision (Bazerman 1990; Wardhaugh and Fuller 2021). Language diversity, hence, can elicit more subtle forms of discrimination compared with the other diversities (Roberts et al 2014). Its unique position would be suitable as a pivot to understand and synthesize the diversity-related socio-technical phenomena in GSD practices. Third, language diversity could form the intersections of diversities with other types of diversities. Individuals with intersections of diversities are likely to be in more disadvantaged situations (see **P15**'s case). For example, female developers from non-English-speaking countries probably need to deal with more challenges. Finally, technological progress, especially in natural language processing, might incur further dynamics in GSD teams by offering more language capital to non-native English speakers since they could eliminate certain language proficiency gaps in text-based communication in GSD teams.

### 7.2 Practical Implications

Our study also has immediate practical implications for practices in GSD teams. First, non-native English speakers should be encouraged to have positive attitudes toward their own language capabilities. They should not feel embarrassed for having an accent, lacking certain social knowledge, etc. Being positive about their language capabilities also reduces their communication avoidance behaviors when interacting with native speakers. Second, native English speakers can be made aware of and acknowledge the challenges and difficulties faced by peer GSD team members who are non-native speakers in acquiring sufficient language capital. Many native speakers have not yet realized they are privileged to be native speakers in GSD teams. They might need to take another perspective to learn what it feels like to work in a foreign language on a daily basis and avoid judging other people based on the native speaker status that is automatically granted to them. Moreover, practitioners may need to be open to modern language technologies such as AI-based machine translations. Negative attributions to the use of such technologies shall be avoided (Mieczkowski 2022), i.e., using such tools shall be a part of the habitus in teams.

Besides, GSD organizations and team leaders should understand non-native English speakers' negative experiences and struggles with language-related issues. They shall publicly promote and endorse language diversity and cultivate proper organizational culture to set the tone. They may explore how to deflate the value of language capital by revising the rules in their organizations. For example, placing competent GSD developers who are non-native English speakers into positions of power. Doing so would also set role models for non-native speakers and encourage them to take their individual agency to improve their disadvantaged positions. They may also legitimize and encourage the use of modern machine translation tools, which may reduce the difficulties non-native speakers face in accessing and accumulating language capital.

## 7.3 Threats to Validity

As with any empirical study, while providing valuable insights for understanding the experiences of GSD developers who were non-native English speakers, our study is not free of threats to validity. We briefly discuss the study's *descriptive validity*, *interpretive validity*, *generalizability*, *evaluative validity*, and *theoretical validity*, respectively (Maxwell 1992).

From the perspective of descriptive validity, we faithfully report the descriptive information extracted from our participants' narratives which were transcribed without any omission. Thus, the potential threats to descriptive validity are minimal. However, an individual's experiences are dynamic by its nature. A single session of interviews with a participant may not fully capture one's dynamic experiences in GSD teams.

From the perspective of interpretive validity, we try our best to accurately understand our participants' viewpoints, thoughts, intentions, and experiences. All authors share the same cultural background with the participants and speak the same native language, which helps us minimize the risk of "validity as culture" where researchers reflect and impose their own cultural point of view for the "others" (Altheide and Johnson 1994). The non-verbal clues that appeared in the interviews were documented, thus enabling us to use these clues to cross-examine participants' intentions. Our rigorous coding process, e.g., having two researchers code independently and resolving conflicting codes collectively, also helps to mitigate the potential threats to interpretive validity.

For qualitative studies, generalizability is more about the transferability of the results rather than universal applicability because such studies are usually concerned with the concepts and characteristics of a select group and its environment (Auerbach and Silverstein 2003). Our study was country-specific because it was performed in China. While we expect similar findings in other countries whose main populations are not English speakers, particularly those sharing similar environments and cultures with China, our findings might not be transferable to many other linguistically and culturally different countries such as India or Latin American countries. These countries have large populations of developers who are non-native English speakers working in GSD teams (Rodeghero et al. 2021). Future research is encouraged to replicate the study design in other countries to develop comprehensive understandings at a global scale.

From the perspective of evaluative validity, we must admit that our own understanding of our participants' situations is unavoidable. Anyway, the problem of how one evaluates the data one receives may raise questions regardless of how well-grounded the evaluation is in the data. It is exactly what drives us to continuously pursue better research design to get us closer to the truth.

Finally, regarding the theoretical validity, we are confident that there is no severe threat to it, partially due to the application of Bourdieu's theoretical framework. We reused its core concepts, which are well-established in a huge body of literature. The data grounded the extensions to the concepts well. We developed theorized relationships among the concepts in the context of GSD teams with the population of non-native English speakers. These relationships maintain high coherence in understanding and explaining the phenomena reflected in our participants' narratives, thus fitting the data.

## 8 Concluding Remarks

This paper reported a qualitative study focusing on non-native English speakers' experiences as a minority group in GSD teams. Using Bourdieu's capital-field-habitus framework, this study reveals the inter-convertible, multifaceted nature of different forms of capital (language, social, symbolic, and economic) in the experiences of GSD developers who were non-native English speakers. We found that non-native speakers' insufficient language capital played an essential role in prohibiting them from accessing and accumulating other capital, thus reproducing the sustained and systematic disadvantaged positions of non-native English speakers in GSD teams. This article thus contributes to the research efforts to develop a holistic and multi-level understanding of non-native English speakers' experiences in GSD teams. Our work demonstrated the necessity of investigating language issues in GSD practices. Therefore, we urge GSD researchers to pay more attention to understanding and effectively addressing language diversity in the increasingly globalized, multilingual, and multicultural software development industry. In the future, we plan to continue this stream of work by utilizing a longitudinal method to uncover the long-term dynamics of individual developers' experience, as well as conducting inductive studies based on more empirical evidence to confirm our findings and explanations. Moreover, while this article focuses mostly on individual experience, many other factors, such as coordination, communication medium, and teamwork quality, could all be associated with language issues. Future studies focusing on these factors, with(out) Bourdieu's framework, must be performed.

## Data Availability

Our data come from interviews with participants working for professional software development organizations. Even with de-identification techniques (removal/redaction of proper names, specific details, etc.), the anonymity of participants and their organizations cannot be fully guaranteed. Due to these privacy and potential legal concerns, we can only share this dataset with others who agree to maintain the anonymity of the participants and obtain their own IRB approval (from their own institutions), but cannot post the dataset in an open repository. Researchers interested in obtaining the dataset should contact the authors directly.